# Artificial Intelligence Strategies for National Security and Safety Standards

Erik Blasch[1], James Sung[2], Tao Nguyen[3],
Chandra Pauline Daniel[4], Alisa Paige Mason[5]

[1]Air Force Office of Scientific Research, erik.blasch.1@us.af.mil
[2]Department of Homeland Security, james.sung@hq.dhs.gov
[3]Defense Intelligence Agency, tao.nguyen@dodiis.mil
[4]National Black Leadership Commission on AIDS cdaniel@hcforensics.org
[5]Guidepost Solutions, pmason@guidepostsolutions.com

**Abstract**

Recent advances in artificial intelligence (AI) have lead to an explosion of multimedia applications (e.g., computer vision (CV) and natural language processing (NLP)) for different domains such as commercial, industrial, and intelligence. In particular, the use of AI applications in a national security environment is often problematic because the opaque nature of the systems leads to an inability for a human to understand how the results came about. A reliance on "black boxes" to generate predictions and inform decisions is potentially disastrous. This paper explores how the application of standards during each stage of the development of an AI system deployed and used in a national security environment would help enable trust. Specifically, we focus on the standards outlined in *Intelligence Community Directive 203* (Analytic Standards) to subject machine outputs to the same rigorous standards as analysis performed by humans.

## 1 Introduction

The current preference and interest in Artificial Intelligence (AI) techniques has a variety of methods for model development to support decision making and predictions. AI includes machine learning (ML) and data, sensor and information fusion [1, 2]. However, for deployment and user understanding and trust, two things are that are needed, and not well addressed are: (1) verification and validation (V&V), and (2) operations and monitoring (O&M); as shown in Fig. 1. Both of these stages help to address questions and concerns from practitioners and users on issues of trust, explainability, robustness, and effectiveness of the AI methods.

The future deployment of AI systems will require standards (or best practices) to gain widespread acceptance of AI methods. For example, what are the standards (e.g., data sheets, forms) that support each stage towards informing the user for understanding (e.g., based on performance – *trust*, based on explainability - *transparency*)? Inherent in the discussion of AI deployment are trust metrics [3, 4] that involve security; while transparency supports safety [5]. Within the Intelligence Community (IC), supporting national security decision makers, analysis aided by AI will need to engender trust, requiring transparency and explainability at each stage of AI deployment testing.

AI is planned for applications such as natural language processing (NLP), computer vision (CV) [6], and data transmission [7]. These examples utilize AI to support a decision maker in their tasks, especially for situations involving huge, multimodal, and novel situations [8].

Product development incorporates many elements including design, test and evaluation, and operational relevance. Typically, a technical community sees an explosion of methods, determines a set of prominent ideas, and develops a set of standards (e.g., guidelines, best practices, metrics, policies, mandates). Current AI methods, such as NLP, have reached the need of developing a set of standards for product development and deployment. One current framework that could be used to guide the development of AI standards in the national security space is the *Intelligence Community Directive 203* [9].

Three assumptions to consider for AI standards include: initial discussion, reference comparison, and community adoption. *Intelligence Community Directive (ICD)* ICD203, entitled "*Analytic Standards*", highlights the multiple data preparation stages inherent for AI methods similar to equipment testing. The second concern is a common reference data set and metrics similar to industry standardized IT8 color charts. Finally, the community must determine what standards are needed based on costs/benefits (e.g. JPEG image compression standard).

To bring the AI community closer to deployment of trustworthy products for national security applications requires common practices for data collection and processing, algorithm and model development, verification and validation (V&V) as well as operations and monitoring (O&M). A standard set of *product labels* would help users better understand elements of AI support.

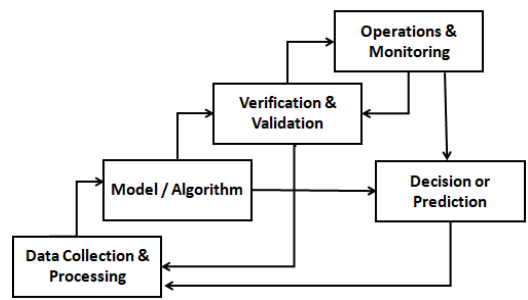

Fig. 1. AI deployment from collections to decisions

This paper reviews the motivation for this effort in Section 2. Section 3 discusses the AI Development Process and Section 4 the need for standards. Section 5 highlights elements of V&V, while Section 6 presents ideas for operations and monitoring. Section 7 recommends the notional AI product label. Section 8 draws conclusions.

## 2 Motivation

As AI applications grow in sophistication, they are likely to become more and more prevalent in the national security space. Natural language processing (NLP), a quickly developing component of AI, utilizes machines to analyze, understand, and generate natural language to:

- Manage the Flood of Information: The amount of unstructured text data generated daily is exponentially increasing. This presents challenges for analysts of various types to classify, triage, and examine all relevant information for their specific problem area of interest, with potential applications to multiple languages. To keep pace with this ever-increasing amount of information, IARPA's BETTER program seeks to develop new tools to enable personalized extraction of semantic information from text for triage and retrieval [10].
- Employ and Create New Means of Generating Insights. AI tackles complex issues [11]. For example, work from MIT researchers in 2019 evaluated an automated fake-news detection system, revealing how machine-learning models catch subtle but consistent differences in the language of factual and false stories [12]. Facebook, in 2017, experimented with AI to understand text that might be advocating for terrorism. They are experimenting with analyzing text that they have already removed for praising or supporting terrorist organizations so they can develop text-based signals that such content may be terrorist propaganda [13].

Use of AI applications in national security environment, however, is often problematic because the opaque nature of the systems leads to an inability for a human to understand how the results came about. Reliance on "black boxes" to generate predictions and inform decisions is potentially disastrous. In this paper, we propose that providing the developer and/or user of a NLP system with standardized information that focuses on the principles of good analysis adopted by IC can help engender greater trust in AI systems used in security.

The ability to trust a machine's output is central to the continued development of beneficial systems. The February 2019 *Executive Order on Maintaining American Leadership in Artificial Intelligence* highlights trust as one of the five guiding principles for future AI development. Section 1(d) states, "The United States must foster public trust and confidence in AI technologies and protect civil liberties, privacy, and American values in their application in order to fully realize the potential of AI technologies for the American people." [14]

In the aftermath of the September 11, 2001 Congress enacted the Intelligence Reform and Terrorism Prevention Act (IRPTA). The Act mandated that the Office of the Director of National Intelligence (ODNI) "assign an individual or entity to be responsible for ensuring that finished intelligence products produced by the intelligence community are timely, objective, independent of political considerations, based on all sources of available intelligence, and employ the standards of proper analytic tradecraft." [15]

ODNI has further codified these standards in *ICD203*. These analytics standards govern the production and evaluation of analytic products; articulate the responsibility of intelligence analysts to strive for excellence, integrity and rigor in their analytic thinking and work [16, 17]. The driving principle of ICD 203 is to engender trust in the IC's analysis. The elements for building this trust include ensuring analytic integrity, rigor, objectivity, relevance, accuracy, timeliness, and assurance for privacy while guarding against bias and politicization. To be tradecraft compliant, AI enabled analysis must also reflect the standards.

## 3 AI Development Process

There are many developments in AI that have emerged over which require standards of processing, deployment, and use. A current focus on AI includes the discussion of *trust and transparency* of mechanisms. For the assessment of AI methods, many metrics are available for the development while the operational performance has yet to be standardized. Furthermore, if the AI tools support prediction, there is a need to better understand the reasoning strategy. A key example is the use of NLP tools for intelligence and security. The "AI" process can generally be gleamed from the "Data Science Hierarchy of needs" as shown in Fig. 2. The hierarchy follows a set of stages.

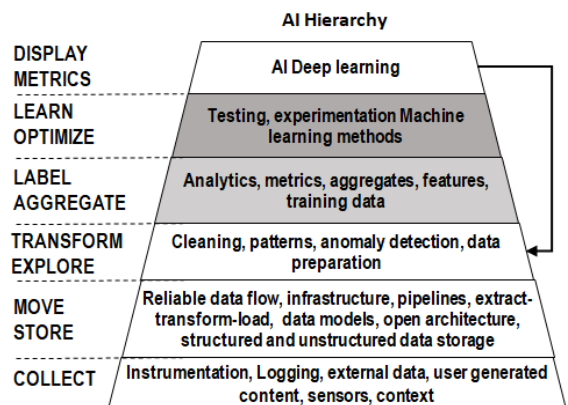

Fig. 2. Data Science Hierarchy of Needs

A key element is that in each of the stages, users are required in the analysis. The AI hierarchy follows a data management flow consisting of (DRUMTC):

- **Data in Collect:** data being sensed
- **Data in Transit:** data in the pipeline being sent
- **Data in Motion:** data dynamically machine processed
- **Data in Use:** data user deems relevant (e.g., labeled)
- **Data at Rest**: data statistically being learned
- **Data in Display:** data results and statistics

Each of these stages could have a variety of mandates, policies or guidelines that determine the process needs. One example that gets overlooked is that there is a balance between *need to know* and *need to share*. As the data is available for AI data analytics, many times the numerous choices for the aggregation, learning, and fusing of data is not available. For example, security, proprietary, and cost of data limits what information is revealed for aggregation.

Also inherent in the challenge is the data attributes that include:

- **Visible:** Who knows the data?
- **Accessible:** Where is the data and is it connected?
- **Understandable**: What does the data represent?
- **Linked:** Why is the active data proposed sharable?
- **Trustworthy:** When can data drive decisions?

### 3.1 Human-Machine Teaming (HMT)

The user would like a trustable result with high reliability, confidence, and credibility – which also render low uncertainty. The problem with many AI/ML techniques, which seek a classification boundary, is that they do not perform uncertainty quantification. Probabilistic approaches (e.g., Bayesian filter) support data in motion (DIM) for which the immediate real-time control needs satisfies reducing uncertainty by minimizing the error. If the data is from multiple sources, perspectives, and situations; then the uncertainty can be further reduced. A control paradigm for data in use (DIU) is the observe-orient-decide-act (OODA) loop. While the OODA loop is well known (Fig. 3), the goal is to have the human and machine team in the learning process. The human seeks results from the machine (observe) to decide and act, which is currently known as *Data-to-Decisions*. From the human's observation of the situation, they would rather select data in concert with the machine decision boundaries – *Decisions-to-Data*. Hence, the decision and data selection should be a combination between machines and humans. Data-to-decisions (classification learning) from the machine should be coordinated with the human that requires decisions-to-data (command and control data collection). Such an example is a user query for text recognition. Together, human-machine AI would help to enhance the usability of current AI systems.

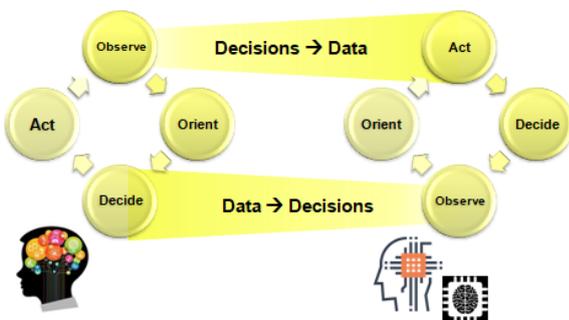

Fig. 3. Human-Machine teaming in OODA loops.

Some representative concerns for the processing of the data for HMT include:

- **Data at Rest**: Provide *structure* (i.e., translations) between data for integration, analysis, and storage;
- **Data in Collect**: Leverage the power of *modeling* from which data is analyzed for information, delivered as knowledge, and supports prediction of future data needs;
- **Data in Transit**: Develop *Data as a Service* architecture that incorporates contextual information, metadata, and information registration to support systems-of-systems design;
- **Data in Motion**: Utilize *feedback control loops* to dynamically adapt to changing priorities, timescales, and mission scenarios; and,
- **Data in Use**: Afford context-based *HMT* based on dynamic mission priorities, users, and situations to balance needs, recommendations, and availabilities.

The data attributes also can leverage emerging concepts:

- **Visible** – Maintaining the User Defined Operating Picture (UDOP) visualization of each domain [18];
- **Accessible** – Providing the compressed information to distributed locations [19];
- **Understandable** – Informing decision-making with context (other domain with historical/planned data collection) [20];
- **Linked** – Developing multi-domain command and control (MDC2) for surveillance and sensing [21];
- **Trusted** – Distributing data in coordination and control that is vetted by others and machines [22].

Given data properties (DRUMCT) and attributes (VAULT), one emerging AI technique that is widely available is natural language processing.

### 3.2 Natural Language Processing

There are many developments in NLP that have emerged over a variety of areas that include [23]:

Entity Analysis:
- Character Recognition
- Text Classification and Categorization
- Named entity recognition

Relationship determination:
- Part-of-Speech Tagging
- Machine Translation
- Speech Recognition

Event Processing:
- Semantic Parsing and Question Answering
- Paraphrase Detection
- Language Generation
- Multi-document Summarization

Each of these NLP applications has a different set of data corpus, decision-making support, as well as measures of importance. The applications follow the NLP pipeline of entity analysis, relationship determination, and event processing. At the simplest level is determining the language and keywords (e.g., *entities*) from documents or handwriting. These techniques support simple applications of doing a search for relevant documents that the user is pointed to for further analysis. A more complex example is question-answering [24].

The second general category would be for *relationships* that link together information within or among a corpus (e.g., statistical relational learning [25]). Moving from tagging to translation and speech recognition has shown promise; however mistakes are still common. Hence, what are the policies for using such techniques in decision-making? Can a translated quote be a meaning representation of what the original speaker indicated? Hence, there are many upstream opportunities, but downstream implications of the output – if not carefully appreciated.

The last category builds on the previous, which includes determining the syntactic, semantic, and sentiment [26] elements of NLP. For example, multi-document summarization includes many advanced techniques [27] that are not well repeatable (or consistent) such that users should be aware of blindly using the outputs without checking the source information.

In all of these NLP techniques, there is a need for society standards that are agreed upon by academic, industrial and government organizations. The academic community has determined some corpus data sets that support challenge problems which could be used for standard processing. Likewise, some common metrics of the credibility, precision-recall, and F score are used;, but there are not consistent standard methods to determine the maturity of the techniques.

## 4 Standards

There are many standards that exist in various research areas that could provide guidance for the AI-NLP community. Some of these areas include computer vision and cyber security, but the support to the intelligence community requires further delineation.

### 4.2 Standards for Multimedia

As there are many developers, a common data set supports research. For example, the image processing community has a variety of examples, as shown in Fig. 4.

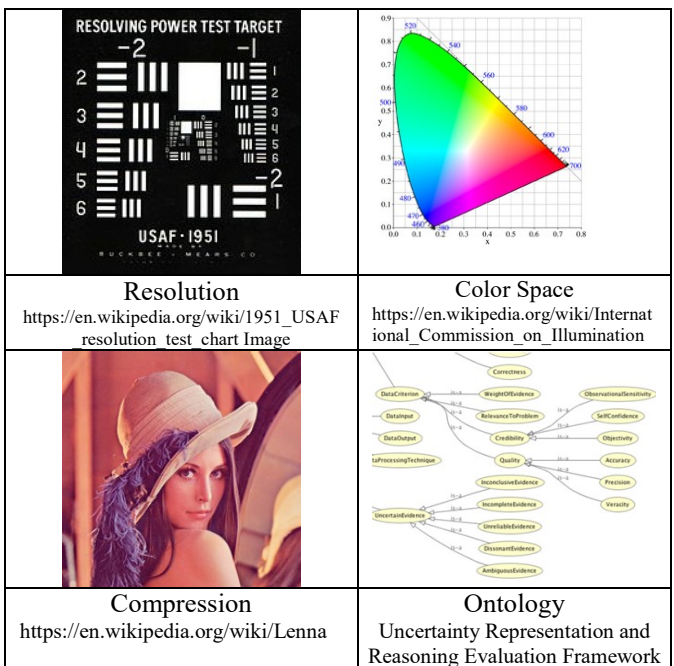

Fig. 4. Multimedia Analytics Standards

As shown in Fig. 4, the *resolution* test chart is used to calibrate the resolution of cameras while the Commission on Illumination created the *color space* chart. Going forward, the computer vision community has utilized a similar image for *compression* analysis. For NLP, an *ontology* provides definitions and delineations of terms, concepts, and meanings applied to words (e.g., entities) that afford a general taxonomy (e.g., relationships).

### 4.1 Standards for Cyber Security

One of the foremost developments in standards is National Institute of Standards, such as for cyber. Other communities have developed security standards for which industry and research follow, as shown in Table 1. The AI community, include image and NLP should determine and adopt a set of standards along the lines of the cyber community including: general, policies, systems, and components. As another example, the avionics or airspace community, through the International Civil Aviation Organization maintains mandates, policies, guidance, and standards for international and interoperable systems.

Table 1 - Industrial Automation and Control Systems (IACS) methods

|  | ISA-62443 |  |
|---|---|---|
| **General** | 1-1 | Terminology, concepts, and models |
|  | 1-2 | Glossary of terms and abbreviations |
|  | 1-3 | System security compliance metrics |
|  | 1-4 | IACS security lifecycle and use case |
| **Policies & Procedures** | 2-1 | Requirements for IACS Security management system (SMS) |
|  | 2-2 | Implementation guidance for IACS SMS |
|  | 2-3 | Patch mgt in IACS environment |
|  | 2-4 | Installation and maintenance requirements for IACS suppliers |
| **System** | 3-1 | Security technologies for IACS |
|  | 3-2 | Security Levels for zones and conduits |
|  | 3-3 | System requirements and levels |
| **Component** | 4-1 | Product development requirements |
|  | 4-2 | Technical security requirements for IACS components |

From: ANSI/ISA 62443 (Formerly ISA-99)- Operational Technical Standards [https://en.wikipedia.org/wiki/Cyber_security_standards]

### 4.3 Standards for Intelligence Community

The intelligence community has a variety of best practices of which the 2015 *Intelligence Community Directive 203* (ICD203), provides a good starting point for AI in the national security environment. A key element is to develop and support a common framework for providing transparency in the analytic process. As with the hierarchy of data analytics, the elements of the ICD203 approximately follow a similar theme with data transformation, aggregation, labeling and display (Fig. 5).

**ICD 203 Data Transformation**

(1) **Properly describes quality and credibility of underlying sources, data, and methodologies**. ICD 206, *Sourcing Requirements/or Disseminated Analytic Products* describes methods to include in source summary statements such as factors affecting source quality and credibility. Such factors can include accuracy and completeness, poss-ible denial and deception, timeliness of information, and technical elements of collection as well as source access, validation, motivation, possible bias, or expertise. Analyst reporting and judgments requires source pedigree and priority, evidence analysis, various assumptions, assessment strengths or weaknesses. Following good sourcing guidance, future AI analytics products should provide standard information on data sources, such as the curation rationale, collection process, labeling choices, data used to train the model and data that had the greatest impact on the model prediction. Other information needs include the intended model use, lineage, and security.

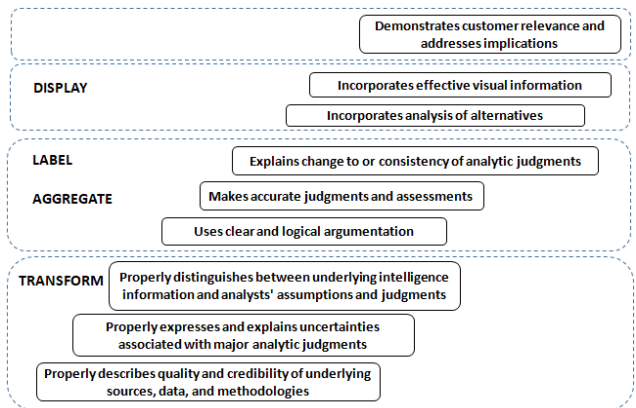

Fig. 5. Elements of Intelligence Community Directive 203.

(2) **Properly expresses and explains uncertainties associated with major analytic judgments**: Analytic products that provide relevant support should indicate data and decision uncertainties. One common method is to support source, analytic, and user reporting. Analysts' confidence is based on individual experience, topic understanding, and quantity and quality of source material. Analytic products should note causes of uncertainty from data, processing, and analysis. One example is the URREF ontology for uncertainty assessment (http://eturwg.c4i.gmu.edu/) [28].

AI analytics could use likelihood or probabilities such as that of STANAG 2511 [29] that correspond to forced choice categories of analysts reporting, (see Table 2).

Table 2 – Reporting Categories

| no chance | very unlikely | unlikely | equal chance | likely | Very likely | almost certain(ly) |
|---|---|---|---|---|---|---|
| remote | highly improbable | improbable | roughly even odds | probable | Highly probable | nearly certain |
| 01 -05% | 05 -20% | 20-45% | 45-55% | 55-80% | 80-95% | 95-99% |

An interesting challenge is to combine the probability of the event, the basis for the assessment, and the confidence in the analysis. Interactive displays and graphical methods would help in the semantic output reporting describing how likely the output is and the confidence level the system has in its judgment. Hence, a standard structured statement similar to that found in intelligence reporting could be "The result YYY is [probability] to occur, based on ABC sources. We have D% confidence in this judgment.."

(3) **Properly distinguishes between underlying intelligence information and analysts' assumptions and judgments**: *Assumptions* are defined as suppositions used to frame or support an argument and affect analytic interpretation. *Judgments* are defined as conclusions based on underlying intelligence information, analysis, and assumptions. Product reports should explicitly state assumptions of an argument or when use to link knowledge gaps; label indicators paramount in judgments; and list various unknowns that would, if known, reduce uncertainty. Documentation can help the user understand when a model is being used in applications that were not represented in the training data and might lead to bad assumptions.

### ICD 203 Data Aggregation and Labeling

(4) **Uses clear and logical argumentation**: Both machines and analyst should use consistent logic (e.g., axioms) in decision making. The use of relevant information, consistent language and syntax, and coherent reasoning would support unambiguous understanding and meaning. Methods for information fusion seek evidential reasoning, plausible explanations, and contradictory results (by way of uncertainty). An effort to make more interpretable models or increase the ability to explain the results post hoc in prose or visualizations can help provide logical justification for a result that is understandable to a human.

(5) **Makes accurate judgments and assessments**: Accuracy in decisions should express precision and recall in spatial, temporal, and modal analysis. The analytics product seeks both relative and absolute solutions between the human and machine interaction. Details on model maintenance to address data drift would allow users understand the model's accuracy. Information on performance metrics, such as accuracy, timeliness, and fairness would assist in judging accuracy.

(6) **Explains change to or consistency of analytic judgments.** Historical assessment comes from social, cultural, and behavioral *modeling*. Modeling patterns and detecting anomalies or changes support the reasoning of events, decisions, and narratives. The results of a model could provide a timeline or history of its analysis on a particular topic to show how it has changed over time.

### ICD 203 Data Display and Decision Relevance

(7) **Incorporates analysis of alternatives.** Most design methods seek course of actions (COAs) through *systematic evaluation* of differing hypotheses to explain events or phenomena, explore results, and imagine possible/plausible futures to mitigate surprise and risk. Machine analytics provide a useful tool to generate many unique assessments and futures of future trends with uncertainty quantification, forecast complexity, and threat impacts. Displaying alternative outcomes and other possible outputs the model considered would help the user better understand how the model could be wrong and prepare for other possibilities.

(8) **Incorporates effective visual information where appropriate**: Reports can use analytics in support of written narratives such as heat maps of spatial *relationships*, timelines of events, and graphs of confidence. Presenting visualizations throughout the development process can be a powerful tool to help the user understand limitations in the training data, how complicated models come to the decisions that they do, and the logic behind model outputs.

(9) **Demonstrates customer relevance and addresses implications**. Relevancy of products and semantic understanding is common in NLP analysis and recently has emerged for visual information. A relevancy assessment provides value by addressing prospects, context, threats, or factors affecting COAs.

Inherent in ICD203 display and relevance is the need for verification and validation (e.g., evaluation) as well as operations and monitoring (e.g., display and reporting).

## 5 Verification and Validation

Various methods are needed to determine the usefulness of the AI-NLP techniques through verification and validation (V&V), shown in Fig. 6.

From one perspective, the definitions include:

- Verification: "Am I measuring the NLP correctly?"
- Validation: "Am I measuring the correct NLP?"

Verification determines whether or not a product, service, or system *complies* with a regulation, requirement, specification, or imposed condition. Verification involves virtual testing with internal customers to meet the technical readiness level (TRL) 4, shown in Table 3. Typical examples of AI-NLP include development testing with known semantic document corpus.

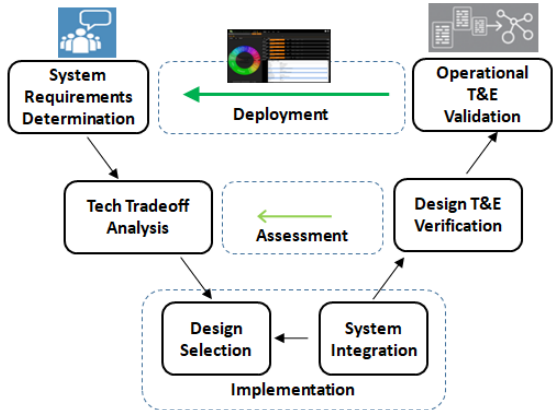

Fig. 6. Verification and Validation

Table 3 – Technology Readiness Levels

| Transition | TRL | Concept |
|---|---|---|
| Deploy | 9 | Evaluated/Certified system in mission operations |
| Test | 8 | Completed system demonstration qualification |
| Operationalize | 7 | Validated prototype in real-world environment |
| Demonstrate | 6 | Validated system adequacy in simulated environment |
| Package | 5 | Verified Technology in real-world environment |
| Develop | 4 | Verified Component in Simulated environment |
| Feasibility | 3 | Demonstrated analytically a Proof-of-Concept |
| Design | 2 | Formulated Technology/Concept/Application |
| Identify | 1 | Observed and Reported Basic Principles |

Validation assures that a product, service, or system *satisfies* the customer or stakeholder requirements. Methods of developing validation include user testing, acceptance and suitability. Validation involves live testing with external customers, typically in a field environment. The results from the AI-NLP must enhance user performance at task flexibility.

There is a need to prepare the systems with first-principle theoretical, learned empirical, or data-augmented dynamical models [30]. One of the challenges is the model interpretability by users and other machines [31]. An evaluation for V&V would determine the bias [32] and the correspondence for human-machine testing for interpretability [33, 34]. Table 4 lists concerns of V&V for NLP evaluation.

Trade studies include modeling and simulation (M&S) over constructive, live, and virtual scenarios [35]. Building from synthetic data and users in a *constructive* simulation, engineering development test and evaluation (DT&E) uses a *virtual* simulation that includes surrogate users with real data. Operational test and evaluation (OT&E) conducts *live* M&S evaluating real people and real data. Beyond efficient solutions to meet real-time needs for measures of performance (MOPs), measures of effectiveness (MOEs) [36] to determine product usefulness.

The ability to convey the methods of effectiveness for different products could be in the form of a food label, such as facts from testing. Ideas that could be listed as facts include the percentages of uncertainty, confidence, completeness, and F score from a standard domain scenario. Building on a food label, the food label ingredients would support the operations and monitoring.

Table 4 – V&V Considerations

|  | Verification | ⟶ | Validation |
|---|---|---|---|
| Testing | Virtual | Constructive | Live |
| Stakeholder | Internal | Product owner | External |
| Process | Data Science | Data analytics | Data prediction |
| Interface | Label | File | Dashboard |
| Models | First-principle | Real-world | Data-augmented |
| TR Level | 4 | 6 | 9 |

## 6 Operations and Monitoring

V&V supports the operations and monitoring (O&M) by way of determining the user support towards the NLP (see Fig. 2), such as data cleaning and labeling. The use of data and tool descriptions, dashboards, and User defined operating pictures (UDOP) supports an effective workflow – going from passive monitoring versus active interaction. A keep attribute of developing systems is a common open standard application programming interface (API).

A key area for O&M is security, privacy [37], and risk mitigation to adversarial attacks. Whether the system is maintained by a server or accessed at the edge [38], there is a need for security. Likewise, the systems must be operational robust and perform even with adversarial attacks [39] that can affect machine training and analysis [40]. However, even adverse examples might be helpful in support of operational maintenance and robustness [41].

Developing a NLP standards board to include the maintenance and consistency of ontologies, terminology, and assessment would ensure products conform to operating needs. The standards would be domain focused such as healthcare, infrastructure, and intelligence. With such standards, effective training on the use of NLP methods could be developed to train a workforce to easily operate on different platforms and tools.

## 7 Recommendations

To build upon previous work and extend it for the national security community, we propose a combination of data statements, dashboards, and information screens throughout the AI development process with a focus on ICD 203 for the IC as a way to subject machine analysis to the same rigorous standards as analysis performed by humans. As of this writing, a number of researchers have proposed strategies for making the AI development process more transparent, understandable, and, in a word, trustworthy. These efforts include:

- Datasheets [42], data statements [43], and "nutrition labels" [44] to better understand the *data* that underlies a machine learning system;
- Cards to clarify the intended use cases of machine learning *algorithms* and minimize their usage in contexts for which they are not well suited [45];
- A "nutritional label" to communicate details of the *ranking methodology* or of the output to the end user [46];
- Factsheets to increase trust in *AI services as a whole* by documenting purpose, performance, safety, security, and provenance for customer examination [47].

As a notional example, Fig. 7 presents an "NLP Attribute Product Label" (NAPL) following the ICD203 guidelines.

Fig. 7. Analytical Label meeting ICD203 guidelines

## 8 Conclusions

The paper presents a need for standards for NLP methods in AI that incorporate V&V and O&M for the deployment of tools to support national security. The main recommendation is to present a standardized data sheet that supports operators in the pragmatic use of various methods following the ICD203 guidelines. Table 5 presents many open questions for an AI NLP system that need to be resolved in future work towards deploying AI systems and standards.

### Acknowledgments

The authors appreciate the support from the Department of Homeland Security (DHS) (Office of Intelligence and Analysis), ODNI Public Private Analytic Exchange Program [AEP-2019]. The views and conclusions contained herein are those of the authors and should not be interpreted as necessarily representing the official policies or endorsements, either expressed or implied, of the United States Air Force, DHS, or DIA. This paper cleared for public release by the U.S. Air Force: 88ABW-2019-4477.

Table 5: NLP AI Questions

| | 1. Data Collection & Processing | 2. Models & Algorithms | 3. Verification & Validation | 4. Operations & Monitoring | 5. Decision & Prediction |
|---|---|---|---|---|---|
| **a. How does it work?** | • What data does the NLP algorithm use for training?<br>• What standard datasets are out there?<br>• What NLP application are the datasets used for? | • How does the algorithm work?<br>• What technique does it use? | • What do we mean by validation?<br>• How do you validate a NLP algorithm?<br>• What efforts are ongoing to verify and validate an algorithm?<br>• What do they measure? | • How robust are NLP systems to adversary attack?<br>• What could someone do? | • What "decisions" are a NLP system making? |
| **b. What are the issues?** | • What issues arise with NLP datasets? | • What types of errors do NLP algorithms suffer from? | • How good are the techniques for V&V of an algorithm?<br>• How does the technique depend on the task? | • What might be the consequences in the context of our use case? | • What issues could arise in the decision making of a NLP system? |
| **c. What standards apply?** | • How can you mitigate these issues?<br>• How might a datasheet for datasets help? | • What standards would help mitigate the errors? | • What standards would apply for V&V of a NLP algorithm? | • What standards would provide users with confidence that the system was robust? | • What would help you trust the decision?<br>• How can you tell if the decision was good or not?<br>• What does it mean for the decision to be interpretable? |
| **d. How to measure or show?** | • What would a user want to know about the dataset? | • How can you determine if the standards are successful? | • What would a user want to know about V&V?<br>• What are we measuring? | • How would you measure the robustness of the NLP system? | • How could you show the user how the decision was made?<br>• What would you need / want / or require the system to convey? |
| **e. What considerations?** | • What are the limitations or unintended consequences of the mitigation? | • What are the considerations or limitations? | • What are the considerations or limitations? | • What considerations or limitations are there to building in adversarial robustness? | • What limitations or considerations would you need to consider? |